\documentclass[aps,prl,reprint]{revtex4-1}
\usepackage{amsmath}
\usepackage{amssymb}
\usepackage{pxfonts}
\usepackage{graphicx}
\usepackage{eucal}
\usepackage{mathrsfs}
\usepackage{theorem}
\usepackage{pifont}
\usepackage{color}
\usepackage{subfig}
\usepackage{cancel}

\newcommand{\brk}[1]{\left( #1 \right)}

\newcommand{\matrixII}[4]{\begin{pmatrix} #1&#2\\#3&#4\end{pmatrix}}

\newcommand{\Exp}[1]{e^{#1}}

\newcommand{\figref}[1]{Fig.~\ref{#1}}

\newcommand{\go}{\bar{\mathfrak{g}}}

\newcommand{\godi}{\go_{\text{dipole}}}

\newcommand{\g}{\mathfrak{g}}

\newcommand{\beq}{\begin{equation}}
\newcommand{\eeq}{\end{equation}}

\newcommand{\burg}{\mathbf{b}}

\begin{document}


\title{Two-dimensional defects in amorphous materials}


\author{Michael Moshe}
\author{Eran Sharon}
\email[]{erans@mail.huji.ac.il}
\author{Ido Levin}
\author{Hillel Aharoni}
\affiliation{Racah Institute of Physics, The Hebrew University, Jerusale}
\author{Raz Kupferman}
\affiliation{Einstein Institute of Mathematics, The Hebrew University, Jerusalem}


\date{\today}

\begin{abstract}
We present a new definition of defects which is based on a Riemannian formulation of incompatible elasticity. Defects are viewed as local deviations of the material's reference metric field, $\go$, from a Euclidian metric. This definition allows the description of defects in amorphous materials and the formulation of the elastic problem, using a single field, $\go$.
We provide a multipole expansion of reference metrics that represent a large family of two-dimensional (2D) localized defects.
The case of a dipole, which corresponds to an edge dislocation is studied analytically, experimentally and numerically.
The quadrupole term, which is studied analytically, as well as higher multipoles of curvature carry local deformations. These multipoles are good candidates for fundamental strain carrying entities in plasticity theories of amorphous materials and for a continuous modeling of recently developed meta-materials.
\end{abstract}

\pacs{61.72.-y}

\maketitle

Defects are known to strongly affect the strength, brittleness and plasticity of solids. In crystalline solids defects appear as intrinsic localized structural deviations of matter from its ordered state. In the continuum approach on the other hand, defects are introduced via \emph{global} constraints on the displacement field \cite{LL}. Therefore, when trying to develop plasticity theories for amorphous materials, in which no structural order exists, one faces the problem of how to intrinsically define defects \cite{kroner,amari,kawaguchi}.

In a solid, a dislocation is characterized by the burgers vector $\burg$, which is a vector-valued measure for the discontinuity in the displacement field.
Specifically, in the presence of a dislocation the total displacement does not vanish along closed loops that surround the dislocation axis. This total displacement is the burgers vector. Another kind of defects are disclinations. Unlike dislocations that are created by translational deformation (linear shift of matter along a plane), disclinations are created by either the removal (positive disclination) or the insertion (negative disclination) of a wedge.
The first attempt to classify dislocations is due to Volterra  \cite{Volterra1907}.
Volterra described elementary states of frustration in elastic materials using cut-shift-weld protocols.
Volterra's constructions provide a list of pathways that result in geometrically frustrated states.
These constructions cannot be regarded as definitions of defects, since very different procedures can result in the same material geometry (see for example  S.1.$a$-$b$).

The calculation of the mechanical state of a solid, and notably the stress field, in the presence of defects is a central task in material science.
A standard approach for calculating the stress field in dislocated bodies is to solve the equilibrium equation of elasticity with the burgers vector as a constraint on the discontinuity of the displacement field.
This yields a well-defined elastic problem, which allows, among other things, to calculate the stress field far from the defect locus, and interactions between dislocations \cite{PK1950,LL}.
In this approach dislocations are external constraints on the continuum model. Their positions and orientations must be determined in advance.

Unlike the case of amorphous materials, in crystals, defects are described as local entities, intrinsic to the material's geometry. For example, on an hexagonal lattice a disclination may appear as a single atom surrounded by 5 (positive disclination) or 7 (negative disclination) neighbors. The hexagonal structure also reveals the connection between dislocations and disclinations: An edge-dislocation is realized by a 5-7  disclinations pair (fig. S.1.$c$) \cite{HirthandLothe,IrvinePleats}. A similar structure can be obtained via Volterra construction. When generated in a thin sheet it results in a cone - anti-cone pair \cite{muller12}.

The local nature of defects in crystals suggests that they can be  described intrinsically by an appropriate material field . Indeed, since the 1950s torsion is used as a measure of the density of dislocations \cite{BergerPanoramic,Kondo,Bilby,kroner}. Recently, the connections between torsion and the dislocation density, as well as the resulting elastic strain were derived \cite{YGdislo,YaBurgers}. It was argued, however, \cite{Wang1968} that a material connection can be defined unambiguously only in the presence of discrete material symmetries. In addition, it was suggested that torsion is a descriptor of disloacations only when an underlying lattice structure exists \cite{kroner, amari, kawaguchi}. It is not clear whether an alternative field theory of defects in amorphous materials can be formulated.



In this work, we model defects within the framework of incompatible elasticity. The description uses riemannian geometry, with no reference to torsion, nor to an underlying material order.
In the framework of incompatible elasticity, an elastic body is modeled as a 3D Riemannian manifold equipped with a ``reference metric'' $\go$,
which represents local equilibrium distances in the material.
Every configuration of the body induces on the manifold a metric, $\g$, which we call the ``actual metric".
The elastic model is fully determined by a constitutive relation, or in the case of a hyper-elastic material, by an energy functional. This energy functional
is an additive measure of local deviations of the actual metric from the reference metric. That is, the elastic energy is of the form,
\[
E = \int W(\g(x); \go(x))\, d\text{Vol}_{\go},
\]
where $d\text{Vol}_{\go}$ is the volume element, and $W$ is a non-negative energy density that vanishes
if and only if $\g(x) = \go(x)$. Incompatibility manifests in that $\g$ cannot be equal to $\go$ everywhere simultaneously.

As stated above, the elastic model is fully captured by the energy functional. In particular, the presence of defects should  also be encoded in the energy functional. Since defects, whether localized or distributed, induce a geometric incompatibility, we expect them to be encoded in the reference metric.

In this paper, we focus on two-dimensional defects in three-dimensional bodies, though much of our analysis can be carried out for three-dimensional defects. Reference metrics that encode 2D defects are
axially symmetric, i.e., they are determined by their value on a cross section.  Incompatibility amounts then to a non-zero Gaussian curvature associated with the reference metric (referred to as the \emph{reference} Gaussian curvature).

Most current literature \cite{Sharon2010,Yael2007,Santangelo2009,efi2,kim2012,Gemmer2011,MarderPapa} studies reference metrics associated with smooth distributions of Gaussian curvature (a discontinuous reference curvature was considered in \cite{PDP2013,SantangeloBistrip}). Such reference metrics are incapable of describing singularities, such as localized defects. Our goal is to identify reference metrics associated with defects.

It is well-known that every 2D metric is locally conformal \cite{BergerPanoramic}, which means that it can be locally expressed as the product of a Euclidean metric and a positive scalar function (the conformal factor). Adopting polar coordinates $(r,\theta)$, we express the metric as follows:
\beq
\go =\Exp{2 \varphi\brk{r,\theta}}\matrixII{1}{0}{0}{r^2}.
\label{eq:conf_go}
\eeq
The function $\Exp{2\varphi\brk{r,\theta}}$ can be interpreted as a local expansion factor.

Since defects are associated with geometric frustration, we model a defect-free amorphous material as a manifold endowed with a Euclidean reference metric. We model an amorphous material with a defect at the origin as a manifold endowed with a reference metric that is locally Euclidean everywhere, except at the origin. Using the \textit{'Brioschi formula'} \cite{doCarmoBaby} the Gaussian curvature corresponding to a metric of the form \eqref{eq:conf_go}  is
\beq
K = \Exp{-2\varphi\brk{r,\theta}} \Delta \varphi\brk{r,\theta},
\eeq
where $\Delta$ is the standard Euclidean Laplacian.
It follows that the reference metric is locally Euclidean if $\varphi$ is harmonic,
\beq
\Delta \varphi\brk{r,\theta} = 0.
\eeq
Using a multipole expansion, we can express a large family of metrics that are locally Euclidean,
\beq
\varphi\brk{r,\theta} = \beta +\alpha \ln r + \sum_{n=1}^{\infty}\brk{A_n r^{n}+ B_n r^{-n}}\brk{\alpha_{n} \cos n \theta +\beta_{n} \sin n \theta},
\label{eq:Multipole}
\eeq
where $\alpha, \beta, \alpha_n, \beta_n, A_n, B_n$ are parameters.
Setting $A_n=0$ amounts to $(r,\theta)$ being standard polar  Euclidean coordinates as $r\to\infty$.

We  now study the geometric interpretation of each multipole term in this expansion by considering an annular domain (which represents a cross section of a punctured 3D cylinder, in which the inner radius serves as a cutoff for the curvature singularity).
The parameter $\beta$ is a homogeneous scaling factor that we may set arbitrarily to zero.
The case in which $\alpha$ is the only non zero coefficient corresponds to a  monopole of Gaussian curvature. It is easy to show (see Supplementary material) that such a metric models a disclination with an excess angle of $2\pi\alpha$. The same geometry can be obtained via a Volterra construction. When applied to thin sheets, these constructions lead to a cone and an e-cone configurations \cite{NS88,Muller2008}.


We next consider the dipole term. Choosing the $\theta=0$ axis parallel to the dipole, the reference metric takes the form
\beq
\godi = e^{{2b \cos \brk{\theta}/r} }
\begin{pmatrix}  1 & 0 \\  0 & r^2 \end{pmatrix}.
\label{eq:EdgeMetric}
\eeq
A dipole is a far field approximation of a pair of monopoles of opposite charge, i.e. the metric induced by a cone anti-cone pair.
In crystals, cone anti-cone pairs (e.g., 5-7 pairs in hexagonal lattices) are edge dislocations.
Thus, the metric \eqref{eq:EdgeMetric} represents an edge dislocation.

A natural question  is how to assign a burgers vector to a reference metric of the form \eqref{eq:EdgeMetric}. Unlike the definition of the burgers vector in conventional elasticity \cite{LL} such a definition should not depend on the actual configuration of the body. In the geometric theory of dislocations in crystalline structures\cite{kroner}, the definition of the burgers vector relies on a notion of parallelism. In crystalline materials, parallel transport of vectors (see \cite{docarmo}) is naturally defined such that the crystalline axes are a parallel frame. In an amorphous material in which the only assumed structure is a metric, the only natural notion of parallelism is induced by the Levi-Civita connection \cite{docarmo}, which is a connection that only reflects the metric structure.

Let $\Pi_q^p$ be the Levi-Civita parallel transport operator from point $q$ to point $p$, induced by \eqref{eq:EdgeMetric} (see SI for details). This operator is path independent since the total curvature enclosed by any loop is zero. Having a well defined parallel transport operator on the reference manifold, we can follow the classical definition of the burgers vector by integrating infinitesimal displacement vectors along a closed loop $\gamma(t)$ on the reference manifold:
\beq
d(\gamma,p) = \int \Pi^{p}_{\gamma\brk{t}} \brk{\dot{\gamma} \brk{t}}\,dt,
\eeq
where $p$ is an arbitrary reference point. It can be rigorously shown \cite{monodromy} that this integral is zero for loops that do not surround the singularity, and it is a constant vector for all loops that surround it once. Moreover, the result is independent of the reference point $p$. Therefore, we can associate a vector with all loops that surround the singularity. This vector, which only relies on the reference metric properties, generalizes the notion of a burgers vector for amorphous materials.

For a reference metric of the form  \eqref{eq:EdgeMetric}, taking $p=(r_0,0)$ and $\gamma$ any loop that surrounds the origin, we find
\beq
\burg \equiv d(\gamma,p) = -2\pi b e^{-\frac{b}{r_0}} \,\hat{y},
\eeq
where $\hat{y}$ is a unit vector along the $y$-axis. The magnitude of this vector (induced by the reference metric) which we identify with the magnitude of the burgers vector, is
\beq
|\burg| =\sqrt{ \langle \burg , \burg \rangle _{\godi} } = 2 \pi b.
\eeq

As expected, the orientation of the burgers vector is perpendicular to the metric dipole.
%

Using Eq.\eqref{eq:EdgeMetric} we experimentally construct a disc with a single dislocation without any cut-shift-weld Volterra procedures. Using an experimental technique similar to that described in \cite{kim2012,Yael2007}, we imposed the  reference metric $\godi$ on an annulus-shaped disc of NIPA gel, by prescribing a varying swelling factor $e^{2\varphi(r,\theta)}$.
The local swelling factor is set by the crosslinking density of the gel, which is determined by the local UV irradiation (see SI).
\begin{figure}[h]
\begin{center}
\includegraphics[width=\columnwidth]{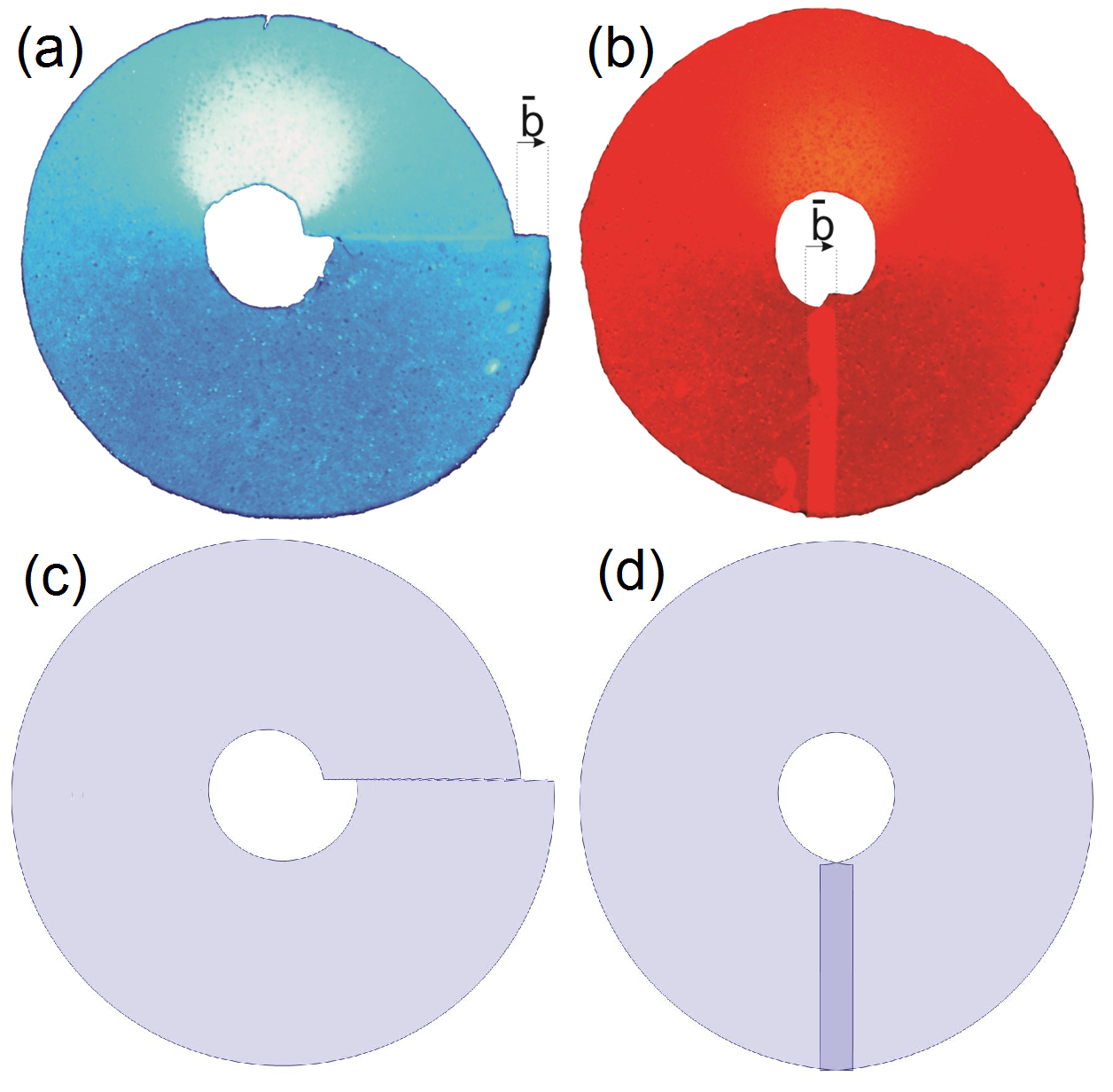}
\end{center}
\caption{Top panel: Experimental realization of a disc with a curvature dipole. NIPA annular gels with a reference metric $\godi$ for $b=5/2\pi$ mm. The sample on the left is cut along the $\theta=0$ axis and the sample on the right is cut along the $\theta=\pi/2$ axis. As a result,  both samples adopt a flat stress-free configuration. In both cases, relaxation is accompanied with a constant shift of about 4 mm. The slightly lighter band in (b) is a region of overlap of two surfaces. Bottom panel: Isometric embedding of the same reference metric as in the experiment, with branch cuts along the rays  $\theta=0$ and  $\theta=\pi/2$.}
  \label{fig:Exp}
\end{figure}

When such an annular disc is flattened, mimicking a cross section of a 3D body, it is residually stressed. The insertion of a radial cut allows the body to relax into a stress-free flat configuration. Figure~\ref{fig:Exp} (top panels) displays two such annuli cut along the $\theta=0$ and $\theta=\frac\pi2$ directions. In both cases, the release of the stress manifests in a deformation typically associated with  edge dislocations: \figref{fig:Exp}$a$ is reminiscent  of a constant discontinuous shift along a planar cut, whereas \figref{fig:Exp}$b$ is reminiscent of the insertion of a half plane.
In both cases the measured displacement of $\sim 4_{mm}$ along the $\theta=\pi/2$ axis is consistent with the burgers vector of magnitude $5_{mm}$, which was prescribed by the reference metric. An analytical calculation (see SI for details) of the stress free state of an annular disc with metric $\godi$ and radial cuts leads to similar configurations (bottom panels of Fig.~\ref{fig:Exp}).
These results show that the geometric structure of the annulus, prior to the insertion of the cut, is the same as the one obtained via Volterra constructions of edge dislocations. Different such cuts would lead to different discontinuous deformations of Volterra type, which are specific realizations of the geometrical object which is encoded in $\godi$.

\begin{figure}[h]
\includegraphics[width=\columnwidth]{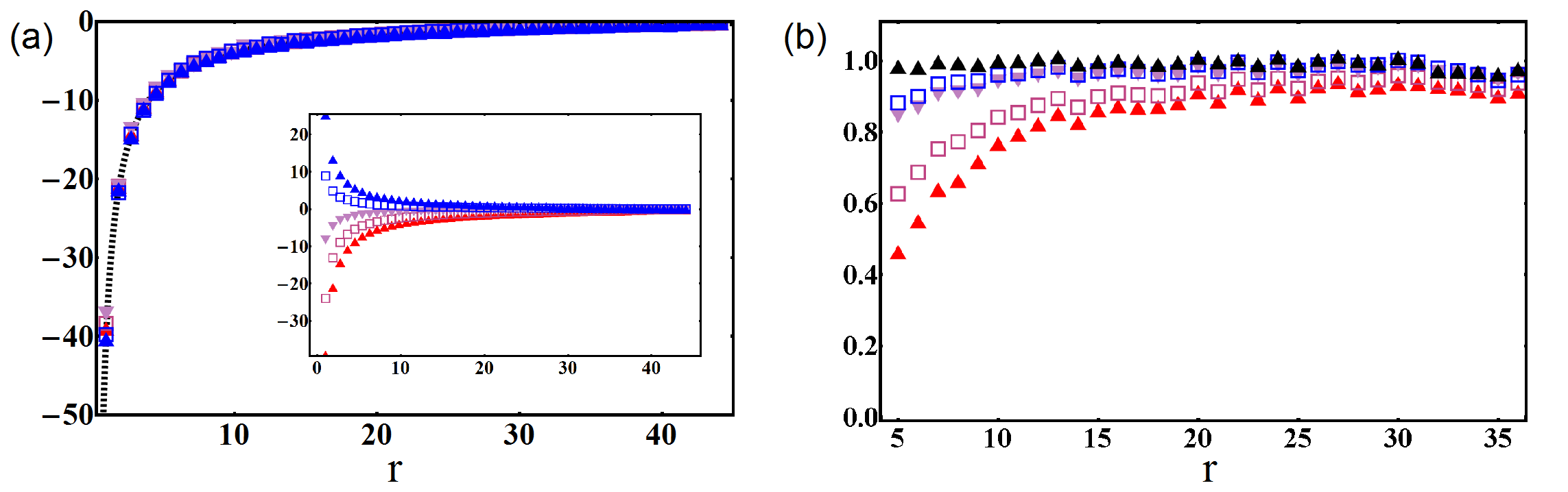}
\caption{
The stress field around a curvature dipole. (a) The stress field $\sigma_{rr}$ as a function of $r$, normalized by $\cos \theta$, for different values of $\theta$. The dotted line is the classical solution for the normalized $\sigma_{rr}$ around an edge dislocation. The inset presents $\sigma_{rr}$ without normalization for $\theta = n \frac{\pi}{7}$ where $n=0-5$ (bottom to top). (b) Dividing $\sigma_{rr}$ computed for increasingly large burgers vectors ($b=0.125,0.4,0.5,0.8,1$ black to red) by the classical solution \cite{HirthandLothe}, we find that this solution become increasingly inaccurate at small distances from the defect.
}
\label{fig:sigma}
\end{figure}


Additional indication for the equivalence between the metric $\godi$ and an edge dislocation can be obtained by comparing the plane stress state in the annulus prior to the insertion of the radial cuts to the classical solution of the stress around a dislocation \cite{HirthandLothe}. We compute the equilibrium planar configuration of the annulus using a finite elements code (see SI for details). For a small burgers vector (or equivalently at large distances) the  solutions coincide (\figref{fig:sigma}$a$). As $\burg$ gets larger we find significant differences between the solutions (\figref{fig:sigma}$b$), where the linearized solution becomes increasingly inaccurate.

\begin{figure}
\begin{center}
\includegraphics[width=\columnwidth]{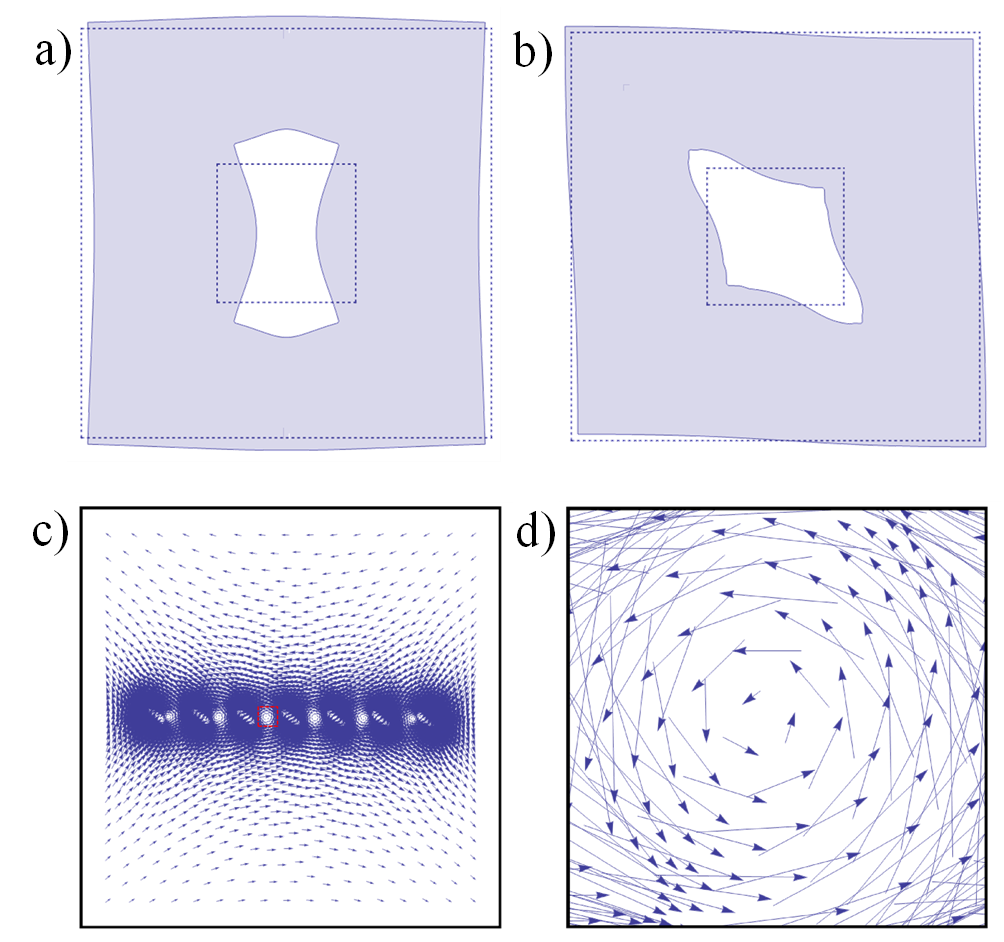}
\end{center}
\caption{Top panel: An analytical calculation of the isometric embedding of reference metrics of quadrupole. The first mode (a), $\varphi_1 = 2 q \cos \theta/r^2$ represent a compression-like deformation while the second mode (b) $\varphi_2 = 2 q \sin \theta/r^2$ represent a shear-like deformation. Both are very similar to the induced deformations in meta-materials. Bottom panel: An analytic solution of the displacement field induced by a linear array of 7 quadrupoles. The pattern of the displacement field, and especially the vortex like patterns between the quadrupoles are very similar to those observed in the collapse of amorphous material due to external shear \cite{Procaccia2} }
\label{fig:Quad}
\end{figure}

The higher multipoles in \eqref{eq:Multipole} correspond to defects that can be generated by localized deformations \cite{monodromy}, similarly to deformations generated by Eshelby inclusions \cite{eshelby}.
Our formalism provides a direct way of computing the deformation associated with such localized defects. For example, an analytical calculation of the two independent modes of deformation that are induced by the quadrupole terms in \eqref{eq:Multipole} are presented in \figref{fig:Quad} (a),(b) (See SI for details of the calculation). Qualitatively similar localized deformations were observed in MD simulations of amorphous materials under remote loading \cite{Procaccia}. It was recently suggested \cite{Procaccia2} that the formation of shear bands during the failure of amorphous solids is generated by the appearance of correlated lines of Eshelby-like singularities. We show analytically and numerically that a linear array of curvature quadrupoles is compatible with a nearly pure shear loading (Fig. S2), a condition which is similar to the external constraint in \cite{Procaccia2}.Moreover, the displacement field in the bulk (\figref{fig:Quad}.c-d), including in the vicinity of the quidrupoles, is also similar to the one observed in \cite{Procaccia2}. In fact, the quadrupole and the higher multipoles of curvature are qualitatively similar to strain carriers in current phenomenological theories of plasticity \cite{STZ,STZ1,STZEshelby}. In our formalism they do not appear as distinct "objects" added to the elastic problem, but they are part of the global reference metric field, which defines the elastic problem.
In addition, there is a strong similarity between the quadrupole induced deformation, and the local elastic deformations observed in meta-materials \cite{Mullin2007,Bertoldi2008,Pedro2010}.

From a computational aspect, our formalism has several clear advantages. It was already shown in (\figref{fig:Exp} and \figref{fig:Quad}) that strain free configurations can be directly computed from the reference metric. In addition, the description of problems that involve more than one defect is a straightforward procedure. The function $\varphi$ in the conformal factor of a reference metric of a body with several separate defects is simply a sum $\varphi = \sum_{i} \varphi_{i}$, where each $\varphi_i$ corresponds to a single defect. This property provides a powerful tool for the description of bodies that contain many defects (see examples in SI), such as the quadrupolar array mentioned above. For problems that involve elastic strains, our formalism is free of geometric linearizations. While this property is of minor importance when considering the far field, it becomes increasingly significant at shorter distances (that can still be much larger than the material dependent cutoff radius). This is well demonstrated in  \figref{fig:sigma}$b$.

In summary, using the geometric approach of incompatible elasticity, we developed a framework for the expression of two-dimensional  defects in three-dimensional amorphous elastic bodies. In this formalism, the body with its defects are described by a single field -- the reference metric field, which is independent of the body's configuration. The formalism, which uses Riemannian geometry, is valid for amorphous materials. We identified a family of defects associated with multipoles of the reference Gaussian curvature and demonstrated the equivalence between the different multipoles and known defects in crystals: The monopole term corresponds to a disclination and the dipole term corresponds to an edge dislocation. Of special interest are the higher multipoles that correspond to defects that can be generated via localized deformations \cite{monodromy}. Such localized deformations are qualitatively similar to local deformations observed in meta-materials as well as to the strain carriers in current phenomenological theories of plasticity \cite{STZ,STZ1,STZEshelby,eshelby}. Combining the new definition and classification of these defects with the computational power of the formalism could be useful in the formulation of future \emph{geometric} continuum theories of plasticity, in which these multipoles are the fundamental strain carrying objects.

\section{}
\subsection{}
\subsubsection{}

\bibliography{Final}

\end{document}